\documentclass[10pt,preprint]{revtex4-1}
\usepackage{graphicx}
\usepackage{color}

\begin{document}
\title{Helicity-selected near-circularly polarized attosecond pulses generated from mixed He-Ne gases}

\author{Chunyang Zhai,$^{1}$}
\email{zhaicy@xynu.edu.cn}
\author{Xiaosong Zhu,$^{2,3}$}
\email{zhuxiaosong@hust.edu.cn}
\author{Yingbin Li,$^{1}$ Qingbin Tang,$^{1}$ Benhai Yu,$^{1}$ Pengfei Lan,$^{2,3}$}
\email{pengfeilan@hust.edu.cn}
\author{Peixiang Lu$^{2,3}$}

\affiliation{
$^1$College of Physics and Electronic Engineering, Xinyang Normal University, Xinyang 464000, China\\
$^2$Wuhan National Laboratory for Optoelectronics and School of Physics, Huazhong University of Science and Technology, Wuhan 430074, China\\
$^3$Hubei Optical Fundamental Research Center, Wuhan 430074, China\\
}

\begin{abstract}
We present and theoretically demonstrate a method for generating helicity-selected near-circularly polarized attosecond pulses in mixed He-Ne gases using bichromatic counter-rotating circularly polarized (BCCP) fields. High-order harmonics driven by BCCP fields exhibit circular polarization for individual orders in the frequency domain, but adjacent orders have opposite helicities. By utilizing the He-Ne mixture, we select only one helical component of the harmonics, resulting in the generation of highly elliptically polarized attosecond pulses in the time domain. Our analyses based on the quantum-orbit theory and the strong field approximation further clarify that the polarization of attosecond pulses is governed by the interference mechanism of high-order harmonics emitted by He and Ne. This combination of BCCP fields and an atomic mixture which requires no alignment in experiments, significantly simplifies the generation of elliptically polarized harmonics dominated by one helical component, thereby paving the way for an efficient and robust method to generate bright attosecond pulses with large ellipticity.
\end{abstract}

\maketitle

Attosecond pulses, which serve as indispensable tools for ultrafast probing, are characterized by short duration, playing an essential role in the field of strong-field physics \cite{Krausz2009}. Presently, the generation of attosecond pulses predominantly relies on the high-order harmonic generation (HHG) in the interaction of femtosecond lasers with matter \cite{Midorikawa2022}. Owing to the continuous development of modulation mechanisms for HHG, the duration of attosecond pulses has steadily decreased since the first demonstration of attosecond pulses in 2001 \cite{250apt,650iap}. The shortest duration of isolated attosecond pulses in the experiment has reached 50 as \cite{43as}. The maximum pulse energy of isolated attosecond pulses has reached the microjoule level \cite{1.3uj}. Besides pulse duration and energy, polarization is another crucial characteristic of light. Recall that the production of high-order harmonics and attosecond pulses is dependent on the laser-induced recollision process \cite{3step}, the efficiency of generating high-order harmonics using elliptically or circularly polarized driving laser fields is notably low \cite{ellIR1}, presenting a persistent challenge in achieving elliptically or circularly polarized high-order harmonics and attosecond pulses. Nonetheless, the importance of elliptically or circularly polarized high-order harmonics and attosecond pulses has been steadily increasing. This is primarily attributed to their potential applications in studying the circular dichroism of magnetic materials, chiral molecules, and electronic spin motion, which have garnered significant attention within the scientific community in recent years \cite{ykj2016}. During the last two decades, scientists have put much effort into polarization control of high-order harmonics both experimentally and theoretically. One type of method involves the use of prefabricated targets, such as aligned molecules or current-carrying orbitals \cite{ali2009,xie2008PRL}. Due to the difficulties in achieving perfect molecular alignment and preparing current-carrying orbitals in experiments, it is challenging to obtain harmonics with a significant ellipticity. Another method involves adjusting the input driving laser fields, such as utilizing bichromatic circularly polarized fields \cite{BCCPxiaolvgao,BCCP-Kfir2015,BCCP119r,BCCPOL,zhuPRL,BCCP115p,PLA2024}, orthogonal two-color laser fields \cite{OTCnc,OTCzhai}, or non-collinear driving fields \cite{Non-collinear2015,Non-collinearChen,NON-Ellis,NON-Han}. Previous studies have demonstrated the advantages of these two-dimensional laser fields in controlling harmonic polarization. Especially in bichromatic counterrotating circularly polarized (BCCP) fields, where the intensities of the fundamental field and its second harmonic (SH) field are comparable, the generated harmonics exhibit efficiency similar to those driven by linearly polarized fields. However, adjacent-order harmonics have almost the same intensity but opposite helicity, which hampers the generation of attosecond pulses with large ellipticity. To generate highly elliptically polarized attosecond pulses, a significant intensity asymmetry must be induced between the right- and left-circularly polarized (RCP and LCP) components of harmonics. While it is possible to selectively enhance one helicity of harmonics by changing the laser intensity ratio of the BCCP field, it results in a reduction of harmonic yield \cite{BCCP119r,BCCPOL,PRLLL}. Besides, the non-collinear scheme also limits harmonic efficiency due to the restricted interaction region. Therefore, achieving the desired ellipticity necessitates sacrificing harmonic efficiency, and vice versa. Recent studies have demonstrated that the interference between harmonics generated from mixed atomic and molecular gases can effectively select a specific helicity, thereby achieving high ellipticity without compromising efficiency \cite{mixedZhai}. To obtain mutually perpendicular components, previous research still depends on the spatial alignment of molecules, which means that the molecules must be pre-aligned in experiments. There is a pressing need for a simple and robust method to generate highly efficient and elliptically polarized high-order harmonics and attosecond pulses.

In this letter, we demonstrate a method for generating attosecond pulses with large ellipticity from mixed atomic gases. Specifically, this method utilizes the interaction between a BCCP field and a He-Ne mixture with opposite orbital parity. It overcomes the technical challenges of molecular alignment and resolves the conflicting demands of harmonic efficiency and ellipticity. Our analyses and simulations show that this approach can generate one helicity dominant high-order harmonics across a wide spectral range, enabling the synthesis of bright near-circularly polarized attosecond pulses.

\begin{figure}[!b]
	\vspace{-6mm}
	\setlength{\abovecaptionskip}{-1pt}
	\setlength{\belowcaptionskip}{0mm}
	\centering
	\includegraphics[width=85mm]{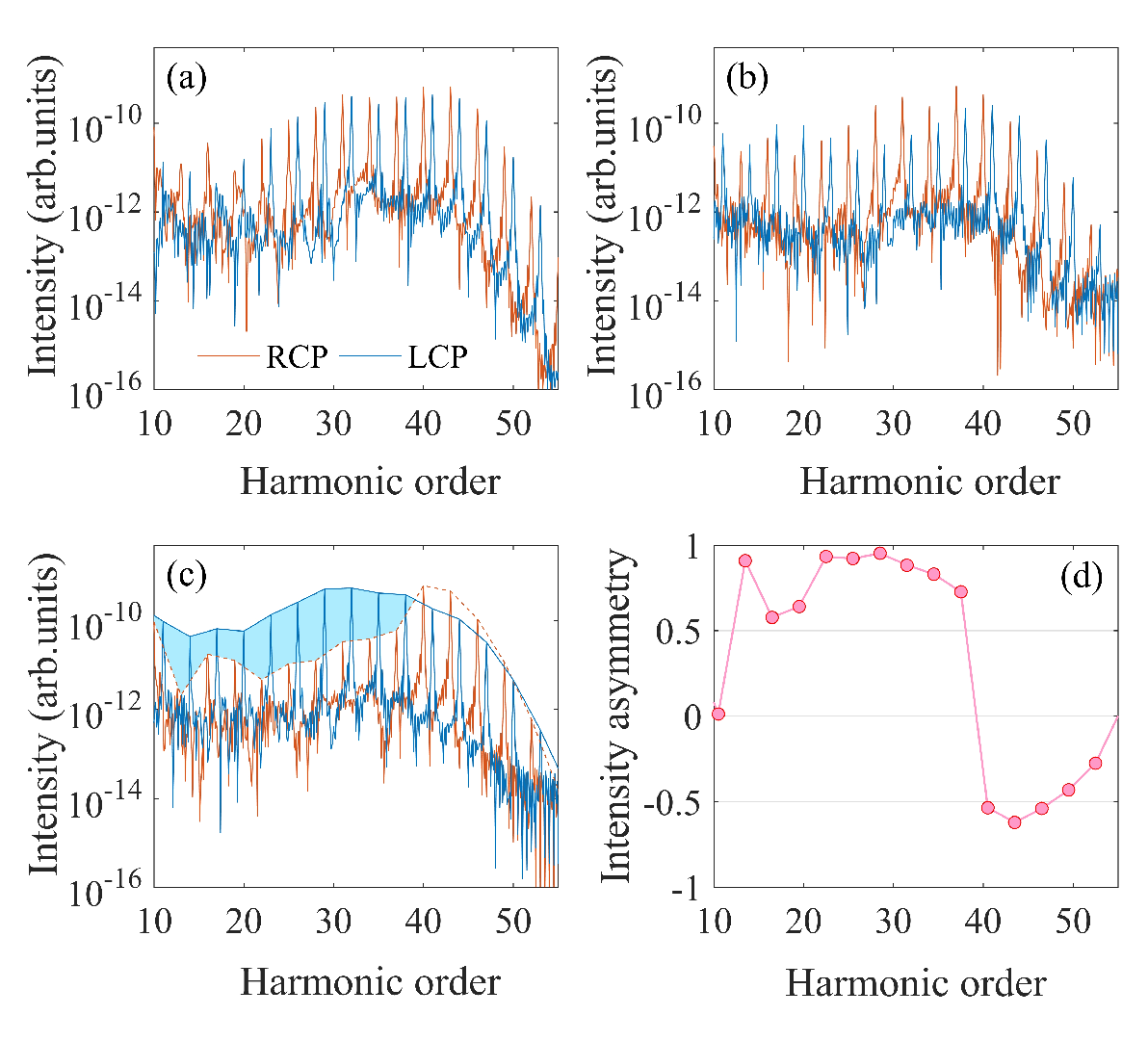}
	\caption{
		High-order harmonic spectra of (a) He and (b) Ne. Colors mark harmonics co-rotating with the fundamental field (brown line) and the SH field (blue line). (c) High-order harmonic spectrum of the He-Ne mixture, exhibiting a complete suppression of RCP in plateau region (shaded area). (d) The intensity asymmetry of harmonic pairs in (c).
	}
	\label{fig1}
\end{figure}

As discussed above, to eliminate the need for pre-aligned molecular targets, it is advisable to use atomic gases in the mixture instead of molecular gases. In this case, two-dimensional driving laser fields are advantageous for obtaining mutually perpendicular harmonic components. In the proof-of-principle study, we employed a BCCP field and a He-Ne mixture. The BCCP field effectively generates and separates RCP and LCP harmonics in the frequency domain. He and Ne have opposite orbital parity and the closest ionization potentials among atomic gases. We first calculate harmonics separately for He and Ne. The laser electric field is $\vec{E}(t)=1/\sqrt2E_0f(t)\{[\text{cos}(\omega t)+\text{cos}(2\omega t+\varphi)]\hat{x}+[\text{sin}(\omega t)-\text{sin}(2\omega t+\varphi)]\hat{y}\}$, which consists of a right circularly polarized fundamental field and a left circularly polarized SH field. $E_0$ and $\omega$ are the amplitude and the angular frequency of the fundamental field, respectively. $\varphi$ represents the relative phase between the fundamental field and its SH field. $f(t)$ is the envelope of the driving laser electric field. In our calculation, the wavelength of the fundamental field is $\lambda=$ 800 nm. $f(t)$ is a trapezoidal envelope that contains a 6-cycle plateau and 2-cycle rising and falling edges (in units of the fundamental field). The intensities of the fundamental and SH fields are $I_\omega=I_{2\omega}=2.4\times10^{14}~\text{W/cm}^2$. Previous experimental research has already demonstrated that the BCCP field with an intensity ratio of 1:1 can generate highly efficient high-order harmonics and attosecond pulses \cite{BCCP119r}. The relative phase $\varphi$ simply rotates the trefoil pattern of the driving field, and $\varphi=0$ is used in our simulation. The ground states of He and Ne, i.e., the 1\textit{s} orbital and the degenerate 2\textit{p} orbitals, are calculated using the Gaussian09 with the B3LYP functional and a 6-31G* basis set \cite{g09}. Based on the strong field approximation (SFA) model \cite{SFA}, the calculated high-order harmonic spectra are shown in Fig. \ref{fig1}, which present the typical shape for the BCCP field approach. Harmonics of orders 3$n$ ($n\in\text{N}$) are forbidden. Harmonics of orders 3$n+$1 and 3$n+$2 exhibit circular polarization and have the same helicities as the fundamental and SH fields, respectively \cite{OCSel,liuxiSel}. Figure \ref{fig1}(a) shows the high-order harmonic spectrum generated in He. The harmonics in pairs [the (3$n+$1)th and the (3$n+$2)th orders] present similar heights. Figure \ref{fig1}(b) shows the high-order harmonic spectrum generated in Ne. One can see that the harmonics present slight intensity asymmetry for the pairs in the plateau region due to the angular momentum of the 2\textit{p} orbital \cite{BCCP115p}.

\begin{figure}[!b]
	\vspace{-6mm}
	\setlength{\abovecaptionskip}{-1pt}
	\setlength{\belowcaptionskip}{0mm}
	\centering
	\includegraphics[width=75mm]{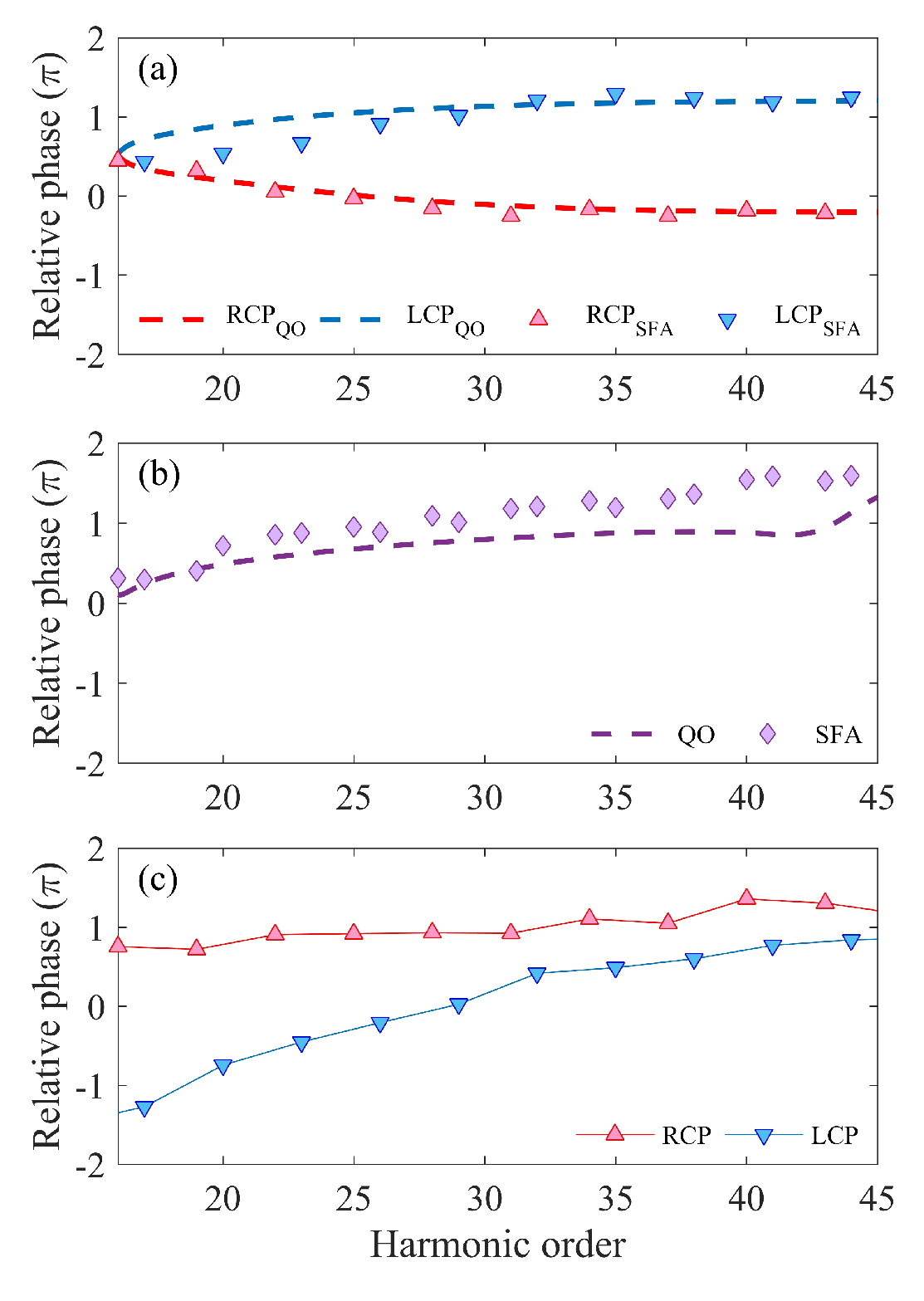}
	\caption{
		(a) The relative phases between	recombination dipole moments of He and Ne for RCP (red dashed line and upward triangles) and LCP (blue dashed line and downward triangles). Results based on the QO analysis are shown as dashed lines and those based on the SFA model are represented by triangles. (b) The purple dashed line and diamonds represent the relative phases acquired during the travel time, obtained from the QO analysis and the SFA model, respectively. (c) The relative phases between high-order harmonic radiation from He and Ne for RCP (upward triangles) and LCP (downward triangles). 
	}
	\label{fig2}
\end{figure}

To obtain significant intensity asymmetry for opposite helicity of harmonics, we next consider mixed gas mediums. The high-order harmonics generated from mixed gases are the result of interference between harmonics emitted by different gas components \cite{mixed2009}. The phase difference between the harmonics radiated from He and Ne is given by \cite{NPlink,rpnature,rpprl}
\begin{equation}\label{E0}
	\Delta\phi\simeq\Delta\phi_i+\Delta\phi_r+\Delta\phi_{\tau}.
\end{equation}
$\Delta\phi_i$ represents the initial relative phase acquired during the ionization process between He and Ne. Previous works have established that $\Delta\phi_i=0$ \cite{rpnature}. $\Delta\phi_r$ is the relative phase between the recombination dipole moments of He and Ne. It depends on the orbital parity of the target, being in phase for one parity and out of phase for the other \cite{mixedZhai}. $\Delta\phi_{\tau}=I_p^{He}\tau^{He}-I_p^{Ne}\tau^{Ne}$ represents the relative phase acquired during the time-interval $\tau$ from ionization to recombination between He and Ne. Here, $I_p^{He/Ne}$ is the ionization potential of He/Ne. $\tau^{He/Ne}$ is the electron travel time in the continuum. It is worth noting that HHG in BCCP fields is dominated by short quantum paths \cite{BCCP115p}. Only short quantum paths will be discussed in the following. For a quantitative explanation, we undertake a quantum-orbit (QO) analysis to elucidate the relative phase between the harmonics emitted by He and Ne \cite{QO,saddlePRL}. In the QO analysis, the ionization and recombination of electrons in the laser field take place in complex time. The complex times $t_i$ and $t_r$ are solutions of the saddle-point equations \cite{saddlePRL}
\begin{equation}\label{E1}
	\textbf{v}^2(t_i)/2+I_p=0,
\end{equation}
\begin{equation}\label{E2}
	\textbf{v}^2(t_r)/2+I_p=q\omega,
\end{equation}
where $q$ is the harmonic order. $\textbf{v}(t)=\textbf{p}+\textbf{A}(t)$ is the instantaneous electron velocity. $\textbf{p}=-\frac{1}{t_r-t_i}\int_{t_i}^{t_r}\textbf{A}(t)\text{d}t$ is the saddle-point momentum and $\textbf{A}(t)=-\int^t\textbf{E}(t')\text{d}t'$ is the vector potential of the laser field. The recombination dipole moment can be expressed as $\textbf{d}(\textbf{v})=\left\langle\Psi|\textbf{r}|\textbf{v}\right\rangle$, where $\Psi$ is the ground state of the target. The relative phase $\Delta\phi_r$ is illustrated in Fig. \ref{fig2}(a). Both the relative phases for the RCP component (red dashed line) and the LCP component (blue dashed line) remain relatively stable in the high energy region above the 20th-order harmonic. The relative phase of RCP is near 0, while that of LCP is close to $\pi$, consistent with the analysis based on the parity of the target orbital \cite{mixedZhai}. The relative phase $\Delta\phi_{\tau}$ is shown in Fig. \ref{fig2}(b) as a purple dashed line. One can see that as the harmonic order increases, the overall trend of the relative phase is generally stable, with only a slight rise.

\begin{figure}[!b]
	\vspace{-6mm}
	\setlength{\abovecaptionskip}{-1pt}
	\centering
	\centering\includegraphics[width=60mm]{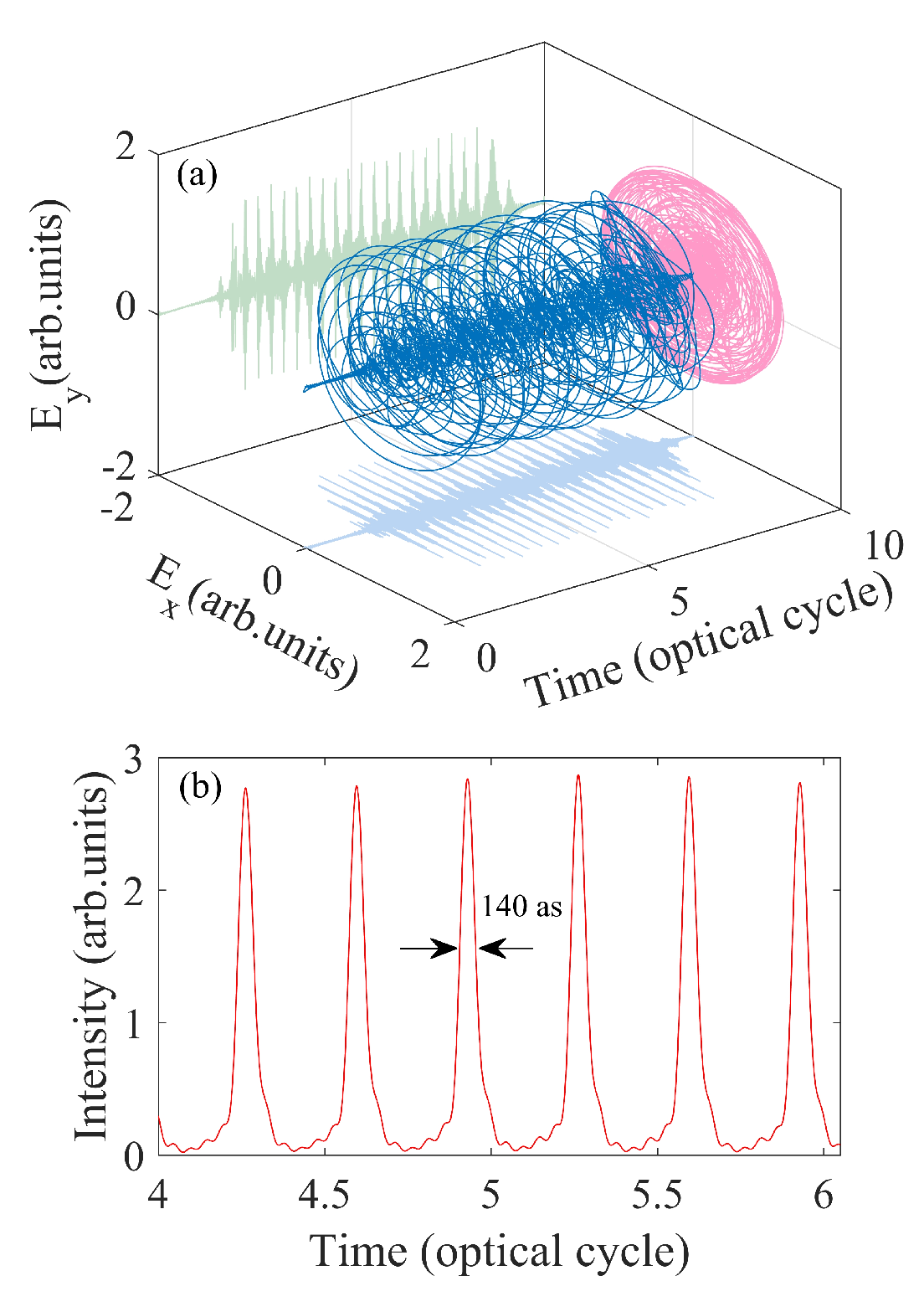}
	\caption{
		A three-dimensional plot of the electric field for the synthesized attosecond pulses with high-order harmonics from mixed gases. (b) The temporal profile of the attosecond pulses.
	}
	\label{fig3}
\end{figure}

To confirm the results of the relative phases, $\Delta\phi_r$ and $\Delta\phi_{\tau}$, obtained in the QO analysis, we further perform the quantum mechanical simulation. In the simulation, we separately confirm $\Delta\phi_r$ and $\Delta\phi_{\tau}$ using the SFA model. First, in order to discuss the relative phase $\Delta\phi_r$, we assume that He and Ne have the same ionization energies. In this case, electrons have an identical travel time, as well as the same phases $\phi_{\tau}$, for the same order harmonic whether they are ionized from 1\textit{s} or 2\textit{p} orbitals. Therefore, the relative phase of high-order harmonics is only contributed by that of the recombination dipole moments. In Fig. \ref{fig2}(a), we show the relative phase $\Delta\phi_r$ as triangles. Both the results of the RCP component (upward triangle) and the LCP component (downward triangle) under the SFA model agree well with the findings of the QO analysis. Second, in order to analyze the relative phase $\Delta\phi_{\tau}$, we assume that He and Ne have the same atomic orbitals, although they have different ionization energies. In this case, the same ground states lead to the same phases $\phi_r$ for the recombination dipole elements. As a result, the relative phase of high-order harmonics is only contributed by that gained during the travel time between two targets. As shown as diamonds in Fig. \ref{fig2}(b), the relative phase $\Delta\phi_{\tau}$ is consistent with the results of the QO analysis.

By combining the results in Fig. \ref{fig2}(a) with those in Fig. \ref{fig2}(b), one can determine the relative phase of the high-order harmonics generated from real He and Ne, i.e., 1\textit{s} orbital for He with $I_p=0.9036$ a.u. and 2\textit{p} orbital for Ne with $I_p=0.7925$ a.u.. In Fig. \ref{fig2}(c), we show the relative phases of high-order harmonics between He and Ne. It is evident that the relative phase of the RCP harmonics is approximately $\pi$, as indicated by the upward triangles in Fig. \ref{fig2}(c). When the harmonic amplitudes contributed by He and Ne become comparable by adjusting the gas mixture ratio, significant destructive interference occurs in the RCP harmonics, as shown in Fig. \ref{fig1}(c). In contrast, the relative phase of the LCP harmonics ranges from -$\pi$/2 to $\pi$/2 between the 20th and the 38th order, resulting in constructive interference in the LCP harmonics. In Fig. \ref{fig1}(c), the shaded area highlights that within a wide spectral range, LCP harmonics dominate the total harmonics in the He-Ne mixed gases. To quantitatively characterize the degree of circular polarization of harmonics, we define the intensity asymmetry of harmonic pairs as $\gamma=(I_{3n+2}-I_{3n+1})/(I_{3n+2}+I_{3n+1})$ \cite{scOE}, where $I_{3n+1}$ and $I_{3n+2}$ represent the intensities of the (3n+1)th and the (3n+2)th order harmonics in each harmonic pair, respectively. As shown in Fig. \ref{fig1}(d), the high-order harmonics exhibit a significant intensity asymmetry close to 1 throughout the entire plateau region. In particular, there is complete suppression for RCP harmonics from the 20th to the 38th order. As a consequence, utilizing harmonics within this spectral range allows for the generation of near-circularly polarized attosecond pulses in the time domain.

Figure 3 shows the output attosecond pulse train, obtained after spectral filtering by selecting the 20th-38th order harmonics presented in Fig. \ref{fig1}(c). The three-dimensional electric field of the attosecond pulse train, the two orthogonal electric field components $E_x$ and $E_y$, and the projection on the $E_x$-$E_y$ plane are depicted in Fig. \ref{fig3}(a). From the helical structures of the electric-field contour plotted in this three-dimensional image, one can directly see that each attosecond pulse is highly elliptically polarized. We evaluate the ellipticity of the generated attosecond pulses by calculating the ratio of the minor axis to the major axis of the elliptically polarized attosecond field, which is about 0.94, very close to that of a circularly polarized pulse. One can see that there are three attosecond pulse radiations per optical cycle as shown in Fig. \ref{fig3}(b). The duration of each attosecond pulse is 140 as.

In conclusion, we investigate the polarization properties of high-order harmonics and attosecond pulses generated from the He-Ne mixture driven by BCCP fields. Both QO analysis and quantum mechanical simulation show that the relative phase of the RCP harmonics radiated from He and Ne remains around $\pi$; in contrast, the relative phase for the LCP harmonics is near zero across a broad spectral range. Then, HHG is dominantly contributed by the LCP harmonics in the broad spectral range, resulting in an attosecond pulse train with an ellipticity of 0.94 and a duration of 140 as. Since the high ellipticity arises from the interference of high-order harmonics emitted by different components, harmonic efficiency will not be affected. Moreover, the use of the atomic gas mixture, without the need for alignment, will significantly streamline the experimental procedure. Our novel approach presents a feasible and simple method for generating highly efficient and elliptically polarized attosecond pulses.

\subsection*{Funding.}
National Natural Science Foundation of China (NSFC) (Grant Nos. 12104389, 12174134, 12074329); Nanhu Scholars Program for Young Scholars of XYNU. The computation is completed in the HPC Platform of Huazhong University of Science and Technology.

\subsection*{Disclosures.}
The authors declare no conflicts of interest.

\subsection*{Data availability.}
Data underlying the results presented in this paper are not publicly available at this time but may be obtained from the authors upon reasonable request.

\bibliography{HeNeRef}

\begin{thebibliography}{37}%
\makeatletter
\providecommand \@ifxundefined [1]{%
 \@ifx{#1\undefined}
}%
\providecommand \@ifnum [1]{%
 \ifnum #1\expandafter \@firstoftwo
 \else \expandafter \@secondoftwo
 \fi
}%
\providecommand \@ifx [1]{%
 \ifx #1\expandafter \@firstoftwo
 \else \expandafter \@secondoftwo
 \fi
}%
\providecommand \natexlab [1]{#1}%
\providecommand \enquote  [1]{``#1''}%
\providecommand \bibnamefont  [1]{#1}%
\providecommand \bibfnamefont [1]{#1}%
\providecommand \citenamefont [1]{#1}%
\providecommand \href@noop [0]{\@secondoftwo}%
\providecommand \href [0]{\begingroup \@sanitize@url \@href}%
\providecommand \@href[1]{\@@startlink{#1}\@@href}%
\providecommand \@@href[1]{\endgroup#1\@@endlink}%
\providecommand \@sanitize@url [0]{\catcode `\\12\catcode `\$12\catcode
  `\&12\catcode `\#12\catcode `\^12\catcode `\_12\catcode `\%12\relax}%
\providecommand \@@startlink[1]{}%
\providecommand \@@endlink[0]{}%
\providecommand \url  [0]{\begingroup\@sanitize@url \@url }%
\providecommand \@url [1]{\endgroup\@href {#1}{\urlprefix }}%
\providecommand \urlprefix  [0]{URL }%
\providecommand \Eprint [0]{\href }%
\providecommand \doibase [0]{http://dx.doi.org/}%
\providecommand \selectlanguage [0]{\@gobble}%
\providecommand \bibinfo  [0]{\@secondoftwo}%
\providecommand \bibfield  [0]{\@secondoftwo}%
\providecommand \translation [1]{[#1]}%
\providecommand \BibitemOpen [0]{}%
\providecommand \bibitemStop [0]{}%
\providecommand \bibitemNoStop [0]{.\EOS\space}%
\providecommand \EOS [0]{\spacefactor3000\relax}%
\providecommand \BibitemShut  [1]{\csname bibitem#1\endcsname}%
\let\auto@bib@innerbib\@empty
\bibitem [{\citenamefont {Krausz}\ and\ \citenamefont
  {Ivanov}(2009)}]{Krausz2009}%
  \BibitemOpen
  \bibfield  {author} {\bibinfo {author} {\bibfnamefont {F.}~\bibnamefont
  {Krausz}}\ and\ \bibinfo {author} {\bibfnamefont {M.}~\bibnamefont
  {Ivanov}},\ }\href {\doibase 10.1103/RevModPhys.81.163} {\bibfield  {journal}
  {\bibinfo  {journal} {Rev. Mod. Phys.}\ }\textbf {\bibinfo {volume} {81}},\
  \bibinfo {pages} {163} (\bibinfo {year} {2009})}\BibitemShut {NoStop}%
\bibitem [{\citenamefont {Midorikawa}(2022)}]{Midorikawa2022}%
  \BibitemOpen
  \bibfield  {author} {\bibinfo {author} {\bibfnamefont {K.}~\bibnamefont
  {Midorikawa}},\ }\href {\doibase 10.1038/s41566-022-00961-9} {\bibfield
  {journal} {\bibinfo  {journal} {Nature Photonics}\ }\textbf {\bibinfo
  {volume} {16}},\ \bibinfo {pages} {267} (\bibinfo {year} {2022})}\BibitemShut
  {NoStop}%
\bibitem [{\citenamefont {Paul}\ \emph {et~al.}(2001)\citenamefont {Paul},
  \citenamefont {Toma}, \citenamefont {Breger}, \citenamefont {Mullot},
  \citenamefont {Augé}, \citenamefont {Balcou}, \citenamefont {Muller},\ and\
  \citenamefont {Agostini}}]{250apt}%
  \BibitemOpen
  \bibfield  {author} {\bibinfo {author} {\bibfnamefont {P.~M.}\ \bibnamefont
  {Paul}}, \bibinfo {author} {\bibfnamefont {E.~S.}\ \bibnamefont {Toma}},
  \bibinfo {author} {\bibfnamefont {P.}~\bibnamefont {Breger}}, \bibinfo
  {author} {\bibfnamefont {G.}~\bibnamefont {Mullot}}, \bibinfo {author}
  {\bibfnamefont {F.}~\bibnamefont {Augé}}, \bibinfo {author} {\bibfnamefont
  {P.}~\bibnamefont {Balcou}}, \bibinfo {author} {\bibfnamefont {H.~G.}\
  \bibnamefont {Muller}}, \ and\ \bibinfo {author} {\bibfnamefont
  {P.}~\bibnamefont {Agostini}},\ }\href {\doibase 10.1126/science.1059413}
  {\bibfield  {journal} {\bibinfo  {journal} {Science}\ }\textbf {\bibinfo
  {volume} {292}},\ \bibinfo {pages} {1689} (\bibinfo {year}
  {2001})}\BibitemShut {NoStop}%
\bibitem [{\citenamefont {Hentschel}\ \emph {et~al.}(2001)\citenamefont
  {Hentschel}, \citenamefont {Kienberger}, \citenamefont {Spielmann},
  \citenamefont {Reider}, \citenamefont {Milosevic}, \citenamefont {Brabec},
  \citenamefont {Corkum}, \citenamefont {Heinzmann}, \citenamefont {Drescher},\
  and\ \citenamefont {Krausz}}]{650iap}%
  \BibitemOpen
  \bibfield  {author} {\bibinfo {author} {\bibfnamefont {M.}~\bibnamefont
  {Hentschel}}, \bibinfo {author} {\bibfnamefont {R.}~\bibnamefont
  {Kienberger}}, \bibinfo {author} {\bibfnamefont {C.}~\bibnamefont
  {Spielmann}}, \bibinfo {author} {\bibfnamefont {G.~A.}\ \bibnamefont
  {Reider}}, \bibinfo {author} {\bibfnamefont {N.}~\bibnamefont {Milosevic}},
  \bibinfo {author} {\bibfnamefont {T.}~\bibnamefont {Brabec}}, \bibinfo
  {author} {\bibfnamefont {P.}~\bibnamefont {Corkum}}, \bibinfo {author}
  {\bibfnamefont {U.}~\bibnamefont {Heinzmann}}, \bibinfo {author}
  {\bibfnamefont {M.}~\bibnamefont {Drescher}}, \ and\ \bibinfo {author}
  {\bibfnamefont {F.}~\bibnamefont {Krausz}},\ }\href {\doibase
  doi:10.1038/35107000} {\bibfield  {journal} {\bibinfo  {journal} {Nature}\
  }\textbf {\bibinfo {volume} {414}},\ \bibinfo {pages} {509} (\bibinfo {year}
  {2001})}\BibitemShut {NoStop}%
\bibitem [{\citenamefont {Gaumnitz}\ \emph {et~al.}(2017)\citenamefont
  {Gaumnitz}, \citenamefont {Jain}, \citenamefont {Pertot}, \citenamefont
  {Huppert}, \citenamefont {Jordan}, \citenamefont {Ardana-Lamas},\ and\
  \citenamefont {W\"{o}rner}}]{43as}%
  \BibitemOpen
  \bibfield  {author} {\bibinfo {author} {\bibfnamefont {T.}~\bibnamefont
  {Gaumnitz}}, \bibinfo {author} {\bibfnamefont {A.}~\bibnamefont {Jain}},
  \bibinfo {author} {\bibfnamefont {Y.}~\bibnamefont {Pertot}}, \bibinfo
  {author} {\bibfnamefont {M.}~\bibnamefont {Huppert}}, \bibinfo {author}
  {\bibfnamefont {I.}~\bibnamefont {Jordan}}, \bibinfo {author} {\bibfnamefont
  {F.}~\bibnamefont {Ardana-Lamas}}, \ and\ \bibinfo {author} {\bibfnamefont
  {H.~J.}\ \bibnamefont {W\"{o}rner}},\ }\href {\doibase 10.1364/OE.25.027506}
  {\bibfield  {journal} {\bibinfo  {journal} {Opt. Express}\ }\textbf {\bibinfo
  {volume} {25}},\ \bibinfo {pages} {27506} (\bibinfo {year}
  {2017})}\BibitemShut {NoStop}%
\bibitem [{\citenamefont {Takahashi}\ \emph {et~al.}(2013)\citenamefont
  {Takahashi}, \citenamefont {Lan}, \citenamefont {Mücke}, \citenamefont
  {Nabekawa},\ and\ \citenamefont {Midorikawa}}]{1.3uj}%
  \BibitemOpen
  \bibfield  {author} {\bibinfo {author} {\bibfnamefont {E.~J.}\ \bibnamefont
  {Takahashi}}, \bibinfo {author} {\bibfnamefont {P.}~\bibnamefont {Lan}},
  \bibinfo {author} {\bibfnamefont {O.~D.}\ \bibnamefont {Mücke}}, \bibinfo
  {author} {\bibfnamefont {Y.}~\bibnamefont {Nabekawa}}, \ and\ \bibinfo
  {author} {\bibfnamefont {K.}~\bibnamefont {Midorikawa}},\ }\href {\doibase
  10.1038/ncomms3691} {\bibfield  {journal} {\bibinfo  {journal} {Nature
  Communications}\ }\textbf {\bibinfo {volume} {4}},\ \bibinfo {pages} {2691}
  (\bibinfo {year} {2013})}\BibitemShut {NoStop}%
\bibitem [{\citenamefont {Corkum}(1993)}]{3step}%
  \BibitemOpen
  \bibfield  {author} {\bibinfo {author} {\bibfnamefont {P.~B.}\ \bibnamefont
  {Corkum}},\ }\href {\doibase 10.1103/PhysRevLett.71.1994} {\bibfield
  {journal} {\bibinfo  {journal} {Phys. Rev. Lett.}\ }\textbf {\bibinfo
  {volume} {71}},\ \bibinfo {pages} {1994} (\bibinfo {year}
  {1993})}\BibitemShut {NoStop}%
\bibitem [{\citenamefont {Budil}\ \emph {et~al.}(1993)\citenamefont {Budil},
  \citenamefont {Sali\`eres}, \citenamefont {L'Huillier}, \citenamefont
  {Ditmire},\ and\ \citenamefont {Perry}}]{ellIR1}%
  \BibitemOpen
  \bibfield  {author} {\bibinfo {author} {\bibfnamefont {K.~S.}\ \bibnamefont
  {Budil}}, \bibinfo {author} {\bibfnamefont {P.}~\bibnamefont {Sali\`eres}},
  \bibinfo {author} {\bibfnamefont {A.}~\bibnamefont {L'Huillier}}, \bibinfo
  {author} {\bibfnamefont {T.}~\bibnamefont {Ditmire}}, \ and\ \bibinfo
  {author} {\bibfnamefont {M.~D.}\ \bibnamefont {Perry}},\ }\href {\doibase
  10.1103/PhysRevA.48.R3437} {\bibfield  {journal} {\bibinfo  {journal} {Phys.
  Rev. A}\ }\textbf {\bibinfo {volume} {48}},\ \bibinfo {pages} {R3437}
  (\bibinfo {year} {1993})}\BibitemShut {NoStop}%
\bibitem [{\citenamefont {Bandrauk}\ and\ \citenamefont
  {Yuan}(2016)}]{ykj2016}%
  \BibitemOpen
  \bibfield  {author} {\bibinfo {author} {\bibfnamefont {A.~D.}\ \bibnamefont
  {Bandrauk}}\ and\ \bibinfo {author} {\bibfnamefont {K.-J.}\ \bibnamefont
  {Yuan}},\ }\href {\doibase 10.1080/00268976.2015.1074742} {\bibfield
  {journal} {\bibinfo  {journal} {Molecular Physics}\ }\textbf {\bibinfo
  {volume} {114}},\ \bibinfo {pages} {344} (\bibinfo {year} {2016})},\ \Eprint
  {http://arxiv.org/abs/https://doi.org/10.1080/00268976.2015.1074742}
  {https://doi.org/10.1080/00268976.2015.1074742} \BibitemShut {NoStop}%
\bibitem [{\citenamefont {Zhou}\ \emph {et~al.}(2009)\citenamefont {Zhou},
  \citenamefont {Lock}, \citenamefont {Wagner}, \citenamefont {Li},
  \citenamefont {Kapteyn},\ and\ \citenamefont {Murnane}}]{ali2009}%
  \BibitemOpen
  \bibfield  {author} {\bibinfo {author} {\bibfnamefont {X.}~\bibnamefont
  {Zhou}}, \bibinfo {author} {\bibfnamefont {R.}~\bibnamefont {Lock}}, \bibinfo
  {author} {\bibfnamefont {N.}~\bibnamefont {Wagner}}, \bibinfo {author}
  {\bibfnamefont {W.}~\bibnamefont {Li}}, \bibinfo {author} {\bibfnamefont
  {H.~C.}\ \bibnamefont {Kapteyn}}, \ and\ \bibinfo {author} {\bibfnamefont
  {M.~M.}\ \bibnamefont {Murnane}},\ }\href {\doibase
  10.1103/PhysRevLett.102.073902} {\bibfield  {journal} {\bibinfo  {journal}
  {Phys. Rev. Lett.}\ }\textbf {\bibinfo {volume} {102}},\ \bibinfo {pages}
  {073902} (\bibinfo {year} {2009})}\BibitemShut {NoStop}%
\bibitem [{\citenamefont {Xie}\ \emph {et~al.}(2008)\citenamefont {Xie},
  \citenamefont {Scrinzi}, \citenamefont {Wickenhauser}, \citenamefont
  {Baltu\ifmmode~\check{s}\else \v{s}\fi{}ka}, \citenamefont {Barth},\ and\
  \citenamefont {Kitzler}}]{xie2008PRL}%
  \BibitemOpen
  \bibfield  {author} {\bibinfo {author} {\bibfnamefont {X.}~\bibnamefont
  {Xie}}, \bibinfo {author} {\bibfnamefont {A.}~\bibnamefont {Scrinzi}},
  \bibinfo {author} {\bibfnamefont {M.}~\bibnamefont {Wickenhauser}}, \bibinfo
  {author} {\bibfnamefont {A.}~\bibnamefont {Baltu\ifmmode~\check{s}\else
  \v{s}\fi{}ka}}, \bibinfo {author} {\bibfnamefont {I.}~\bibnamefont {Barth}},
  \ and\ \bibinfo {author} {\bibfnamefont {M.}~\bibnamefont {Kitzler}},\ }\href
  {\doibase 10.1103/PhysRevLett.101.033901} {\bibfield  {journal} {\bibinfo
  {journal} {Phys. Rev. Lett.}\ }\textbf {\bibinfo {volume} {101}},\ \bibinfo
  {pages} {033901} (\bibinfo {year} {2008})}\BibitemShut {NoStop}%
\bibitem [{\citenamefont {Fleischer}\ \emph {et~al.}(2014)\citenamefont
  {Fleischer}, \citenamefont {Kfir}, \citenamefont {Diskin}, \citenamefont
  {Sidorenko},\ and\ \citenamefont {Cohen}}]{BCCPxiaolvgao}%
  \BibitemOpen
  \bibfield  {author} {\bibinfo {author} {\bibfnamefont {A.}~\bibnamefont
  {Fleischer}}, \bibinfo {author} {\bibfnamefont {O.}~\bibnamefont {Kfir}},
  \bibinfo {author} {\bibfnamefont {T.}~\bibnamefont {Diskin}}, \bibinfo
  {author} {\bibfnamefont {P.}~\bibnamefont {Sidorenko}}, \ and\ \bibinfo
  {author} {\bibfnamefont {O.}~\bibnamefont {Cohen}},\ }\href {\doibase
  10.1038/nphoton.2014.108} {\bibfield  {journal} {\bibinfo  {journal} {Nature
  Photonics}\ }\textbf {\bibinfo {volume} {8}},\ \bibinfo {pages} {543}
  (\bibinfo {year} {2014})}\BibitemShut {NoStop}%
\bibitem [{\citenamefont {Kfir}\ \emph {et~al.}(2015)\citenamefont {Kfir},
  \citenamefont {Grychtol}, \citenamefont {Turgut}, \citenamefont {Knut},
  \citenamefont {Zusin}, \citenamefont {Popmintchev}, \citenamefont
  {Popmintchev}, \citenamefont {Nembach}, \citenamefont {Shaw}, \citenamefont
  {Fleischer}, \citenamefont {Kapteyn}, \citenamefont {Murnane},\ and\
  \citenamefont {Cohen}}]{BCCP-Kfir2015}%
  \BibitemOpen
  \bibfield  {author} {\bibinfo {author} {\bibfnamefont {O.}~\bibnamefont
  {Kfir}}, \bibinfo {author} {\bibfnamefont {P.}~\bibnamefont {Grychtol}},
  \bibinfo {author} {\bibfnamefont {E.}~\bibnamefont {Turgut}}, \bibinfo
  {author} {\bibfnamefont {R.}~\bibnamefont {Knut}}, \bibinfo {author}
  {\bibfnamefont {D.}~\bibnamefont {Zusin}}, \bibinfo {author} {\bibfnamefont
  {D.}~\bibnamefont {Popmintchev}}, \bibinfo {author} {\bibfnamefont
  {T.}~\bibnamefont {Popmintchev}}, \bibinfo {author} {\bibfnamefont
  {H.}~\bibnamefont {Nembach}}, \bibinfo {author} {\bibfnamefont {J.~M.}\
  \bibnamefont {Shaw}}, \bibinfo {author} {\bibfnamefont {A.}~\bibnamefont
  {Fleischer}}, \bibinfo {author} {\bibfnamefont {H.}~\bibnamefont {Kapteyn}},
  \bibinfo {author} {\bibfnamefont {M.}~\bibnamefont {Murnane}}, \ and\
  \bibinfo {author} {\bibfnamefont {O.}~\bibnamefont {Cohen}},\ }\href
  {\doibase 10.1038/nphoton.2014.293} {\bibfield  {journal} {\bibinfo
  {journal} {Nat. Photon.}\ }\textbf {\bibinfo {volume} {9}},\ \bibinfo {pages}
  {99} (\bibinfo {year} {2015})}\BibitemShut {NoStop}%
\bibitem [{\citenamefont {Dorney}\ \emph {et~al.}(2017)\citenamefont {Dorney},
  \citenamefont {Ellis}, \citenamefont {Hern\'andez-Garc\'{\i}a}, \citenamefont
  {Hickstein}, \citenamefont {Mancuso}, \citenamefont {Brooks}, \citenamefont
  {Fan}, \citenamefont {Fan}, \citenamefont {Zusin}, \citenamefont {Gentry},
  \citenamefont {Grychtol}, \citenamefont {Kapteyn},\ and\ \citenamefont
  {Murnane}}]{BCCP119r}%
  \BibitemOpen
  \bibfield  {author} {\bibinfo {author} {\bibfnamefont {K.~M.}\ \bibnamefont
  {Dorney}}, \bibinfo {author} {\bibfnamefont {J.~L.}\ \bibnamefont {Ellis}},
  \bibinfo {author} {\bibfnamefont {C.}~\bibnamefont
  {Hern\'andez-Garc\'{\i}a}}, \bibinfo {author} {\bibfnamefont {D.~D.}\
  \bibnamefont {Hickstein}}, \bibinfo {author} {\bibfnamefont {C.~A.}\
  \bibnamefont {Mancuso}}, \bibinfo {author} {\bibfnamefont {N.}~\bibnamefont
  {Brooks}}, \bibinfo {author} {\bibfnamefont {T.}~\bibnamefont {Fan}},
  \bibinfo {author} {\bibfnamefont {G.}~\bibnamefont {Fan}}, \bibinfo {author}
  {\bibfnamefont {D.}~\bibnamefont {Zusin}}, \bibinfo {author} {\bibfnamefont
  {C.}~\bibnamefont {Gentry}}, \bibinfo {author} {\bibfnamefont
  {P.}~\bibnamefont {Grychtol}}, \bibinfo {author} {\bibfnamefont {H.~C.}\
  \bibnamefont {Kapteyn}}, \ and\ \bibinfo {author} {\bibfnamefont {M.~M.}\
  \bibnamefont {Murnane}},\ }\href {\doibase 10.1103/PhysRevLett.119.063201}
  {\bibfield  {journal} {\bibinfo  {journal} {Phys. Rev. Lett.}\ }\textbf
  {\bibinfo {volume} {119}},\ \bibinfo {pages} {063201} (\bibinfo {year}
  {2017})}\BibitemShut {NoStop}%
\bibitem [{\citenamefont {Zhavoronkov}\ and\ \citenamefont
  {Ivanov}(2017)}]{BCCPOL}%
  \BibitemOpen
  \bibfield  {author} {\bibinfo {author} {\bibfnamefont {N.}~\bibnamefont
  {Zhavoronkov}}\ and\ \bibinfo {author} {\bibfnamefont {M.}~\bibnamefont
  {Ivanov}},\ }\href {\doibase 10.1364/OL.42.004720} {\bibfield  {journal}
  {\bibinfo  {journal} {Opt. Lett.}\ }\textbf {\bibinfo {volume} {42}},\
  \bibinfo {pages} {4720} (\bibinfo {year} {2017})}\BibitemShut {NoStop}%
\bibitem [{\citenamefont {Zhu}\ \emph {et~al.}(2022)\citenamefont {Zhu},
  \citenamefont {Lu},\ and\ \citenamefont {Lein}}]{zhuPRL}%
  \BibitemOpen
  \bibfield  {author} {\bibinfo {author} {\bibfnamefont {X.}~\bibnamefont
  {Zhu}}, \bibinfo {author} {\bibfnamefont {P.}~\bibnamefont {Lu}}, \ and\
  \bibinfo {author} {\bibfnamefont {M.}~\bibnamefont {Lein}},\ }\href {\doibase
  10.1103/PhysRevLett.128.030401} {\bibfield  {journal} {\bibinfo  {journal}
  {Phys. Rev. Lett.}\ }\textbf {\bibinfo {volume} {128}},\ \bibinfo {pages}
  {030401} (\bibinfo {year} {2022})}\BibitemShut {NoStop}%
\bibitem [{\citenamefont {Medi\ifmmode~\check{s}\else \v{s}\fi{}auskas}\ \emph
  {et~al.}(2015)\citenamefont {Medi\ifmmode~\check{s}\else \v{s}\fi{}auskas},
  \citenamefont {Wragg}, \citenamefont {van~der Hart},\ and\ \citenamefont
  {Ivanov}}]{BCCP115p}%
  \BibitemOpen
  \bibfield  {author} {\bibinfo {author} {\bibfnamefont {L.}~\bibnamefont
  {Medi\ifmmode~\check{s}\else \v{s}\fi{}auskas}}, \bibinfo {author}
  {\bibfnamefont {J.}~\bibnamefont {Wragg}}, \bibinfo {author} {\bibfnamefont
  {H.}~\bibnamefont {van~der Hart}}, \ and\ \bibinfo {author} {\bibfnamefont
  {M.~Y.}\ \bibnamefont {Ivanov}},\ }\href {\doibase
  10.1103/PhysRevLett.115.153001} {\bibfield  {journal} {\bibinfo  {journal}
  {Phys. Rev. Lett.}\ }\textbf {\bibinfo {volume} {115}},\ \bibinfo {pages}
  {153001} (\bibinfo {year} {2015})}\BibitemShut {NoStop}%
\bibitem [{\citenamefont {Rajpoot}\ \emph {et~al.}(2024)\citenamefont
  {Rajpoot}, \citenamefont {Holkundkar}, \citenamefont {Rana},\ and\
  \citenamefont {Dixit}}]{PLA2024}%
  \BibitemOpen
  \bibfield  {author} {\bibinfo {author} {\bibfnamefont {R.}~\bibnamefont
  {Rajpoot}}, \bibinfo {author} {\bibfnamefont {A.~R.}\ \bibnamefont
  {Holkundkar}}, \bibinfo {author} {\bibfnamefont {N.}~\bibnamefont {Rana}}, \
  and\ \bibinfo {author} {\bibfnamefont {G.}~\bibnamefont {Dixit}},\ }\href
  {\doibase https://doi.org/10.1016/j.physleta.2023.129241} {\bibfield
  {journal} {\bibinfo  {journal} {Physics Letters A}\ }\textbf {\bibinfo
  {volume} {493}},\ \bibinfo {pages} {129241} (\bibinfo {year}
  {2024})}\BibitemShut {NoStop}%
\bibitem [{\citenamefont {Lambert}\ \emph {et~al.}(2015)\citenamefont
  {Lambert}, \citenamefont {Vodungbo}, \citenamefont {Gautier}, \citenamefont
  {Mahieu}, \citenamefont {Malka}, \citenamefont {Sebban}, \citenamefont
  {Zeitoun}, \citenamefont {Luning}, \citenamefont {Perron}, \citenamefont
  {Andreev}, \citenamefont {Stremoukhov}, \citenamefont {Ardana-Lamas},
  \citenamefont {Dax}, \citenamefont {Hauri}, \citenamefont {Sardinha},\ and\
  \citenamefont {Fajardo}}]{OTCnc}%
  \BibitemOpen
  \bibfield  {author} {\bibinfo {author} {\bibfnamefont {G.}~\bibnamefont
  {Lambert}}, \bibinfo {author} {\bibfnamefont {B.}~\bibnamefont {Vodungbo}},
  \bibinfo {author} {\bibfnamefont {J.}~\bibnamefont {Gautier}}, \bibinfo
  {author} {\bibfnamefont {B.}~\bibnamefont {Mahieu}}, \bibinfo {author}
  {\bibfnamefont {V.}~\bibnamefont {Malka}}, \bibinfo {author} {\bibfnamefont
  {S.}~\bibnamefont {Sebban}}, \bibinfo {author} {\bibfnamefont
  {P.}~\bibnamefont {Zeitoun}}, \bibinfo {author} {\bibfnamefont
  {J.}~\bibnamefont {Luning}}, \bibinfo {author} {\bibfnamefont
  {J.}~\bibnamefont {Perron}}, \bibinfo {author} {\bibfnamefont
  {A.}~\bibnamefont {Andreev}}, \bibinfo {author} {\bibfnamefont
  {S.}~\bibnamefont {Stremoukhov}}, \bibinfo {author} {\bibfnamefont
  {F.}~\bibnamefont {Ardana-Lamas}}, \bibinfo {author} {\bibfnamefont
  {A.}~\bibnamefont {Dax}}, \bibinfo {author} {\bibfnamefont {C.~P.}\
  \bibnamefont {Hauri}}, \bibinfo {author} {\bibfnamefont {A.}~\bibnamefont
  {Sardinha}}, \ and\ \bibinfo {author} {\bibfnamefont {M.}~\bibnamefont
  {Fajardo}},\ }\href {\doibase 10.1038/ncomms7167} {\bibfield  {journal}
  {\bibinfo  {journal} {Nature Communications}\ }\textbf {\bibinfo {volume}
  {6}},\ \bibinfo {pages} {6167} (\bibinfo {year} {2015})}\BibitemShut
  {NoStop}%
\bibitem [{\citenamefont {Zhai}\ \emph {et~al.}(2020)\citenamefont {Zhai},
  \citenamefont {Shao}, \citenamefont {Lan}, \citenamefont {Wang},
  \citenamefont {Zhang}, \citenamefont {Yuan}, \citenamefont {Njoroge},
  \citenamefont {He},\ and\ \citenamefont {Lu}}]{OTCzhai}%
  \BibitemOpen
  \bibfield  {author} {\bibinfo {author} {\bibfnamefont {C.}~\bibnamefont
  {Zhai}}, \bibinfo {author} {\bibfnamefont {R.}~\bibnamefont {Shao}}, \bibinfo
  {author} {\bibfnamefont {P.}~\bibnamefont {Lan}}, \bibinfo {author}
  {\bibfnamefont {B.}~\bibnamefont {Wang}}, \bibinfo {author} {\bibfnamefont
  {Y.}~\bibnamefont {Zhang}}, \bibinfo {author} {\bibfnamefont
  {H.}~\bibnamefont {Yuan}}, \bibinfo {author} {\bibfnamefont {S.~M.}\
  \bibnamefont {Njoroge}}, \bibinfo {author} {\bibfnamefont {L.}~\bibnamefont
  {He}}, \ and\ \bibinfo {author} {\bibfnamefont {P.}~\bibnamefont {Lu}},\
  }\href {\doibase 10.1103/PhysRevA.101.053407} {\bibfield  {journal} {\bibinfo
   {journal} {Phys. Rev. A}\ }\textbf {\bibinfo {volume} {101}},\ \bibinfo
  {pages} {053407} (\bibinfo {year} {2020})}\BibitemShut {NoStop}%
\bibitem [{\citenamefont {Hickstein}\ \emph {et~al.}(2015)\citenamefont
  {Hickstein}, \citenamefont {Dollar}, \citenamefont {Grychtol}, \citenamefont
  {Ellis}, \citenamefont {Knut}, \citenamefont {Hernández-García},
  \citenamefont {Zusin}, \citenamefont {Gentry}, \citenamefont {Shaw},
  \citenamefont {Fan}, \citenamefont {Dorney}, \citenamefont {Becker},
  \citenamefont {Jaroń-Becker}, \citenamefont {Kapteyn}, \citenamefont
  {Murnane},\ and\ \citenamefont {Durfee}}]{Non-collinear2015}%
  \BibitemOpen
  \bibfield  {author} {\bibinfo {author} {\bibfnamefont {D.~D.}\ \bibnamefont
  {Hickstein}}, \bibinfo {author} {\bibfnamefont {F.~J.}\ \bibnamefont
  {Dollar}}, \bibinfo {author} {\bibfnamefont {P.}~\bibnamefont {Grychtol}},
  \bibinfo {author} {\bibfnamefont {J.~L.}\ \bibnamefont {Ellis}}, \bibinfo
  {author} {\bibfnamefont {R.}~\bibnamefont {Knut}}, \bibinfo {author}
  {\bibfnamefont {C.}~\bibnamefont {Hernández-García}}, \bibinfo {author}
  {\bibfnamefont {D.}~\bibnamefont {Zusin}}, \bibinfo {author} {\bibfnamefont
  {C.}~\bibnamefont {Gentry}}, \bibinfo {author} {\bibfnamefont {J.~M.}\
  \bibnamefont {Shaw}}, \bibinfo {author} {\bibfnamefont {T.}~\bibnamefont
  {Fan}}, \bibinfo {author} {\bibfnamefont {K.~M.}\ \bibnamefont {Dorney}},
  \bibinfo {author} {\bibfnamefont {A.}~\bibnamefont {Becker}}, \bibinfo
  {author} {\bibfnamefont {A.}~\bibnamefont {Jaroń-Becker}}, \bibinfo {author}
  {\bibfnamefont {H.~C.}\ \bibnamefont {Kapteyn}}, \bibinfo {author}
  {\bibfnamefont {M.~M.}\ \bibnamefont {Murnane}}, \ and\ \bibinfo {author}
  {\bibfnamefont {C.~G.}\ \bibnamefont {Durfee}},\ }\href {\doibase
  10.1038/nphoton.2015.181} {\bibfield  {journal} {\bibinfo  {journal} {Nature
  Photonics}\ }\textbf {\bibinfo {volume} {9}},\ \bibinfo {pages} {743}
  (\bibinfo {year} {2015})}\BibitemShut {NoStop}%
\bibitem [{\citenamefont {Huang}\ \emph {et~al.}(2018)\citenamefont {Huang},
  \citenamefont {Hernández-García}, \citenamefont {Huang}, \citenamefont
  {Huang}, \citenamefont {Lu}, \citenamefont {Rego}, \citenamefont {Hickstein},
  \citenamefont {Ellis}, \citenamefont {Jaron-Becker}, \citenamefont {Becker},
  \citenamefont {Yang}, \citenamefont {Durfee}, \citenamefont {Plaja},
  \citenamefont {Kapteyn}, \citenamefont {Murnane}, \citenamefont {Kung},\ and\
  \citenamefont {Chen}}]{Non-collinearChen}%
  \BibitemOpen
  \bibfield  {author} {\bibinfo {author} {\bibfnamefont {P.-C.}\ \bibnamefont
  {Huang}}, \bibinfo {author} {\bibfnamefont {C.}~\bibnamefont
  {Hernández-García}}, \bibinfo {author} {\bibfnamefont {J.-T.}\ \bibnamefont
  {Huang}}, \bibinfo {author} {\bibfnamefont {P.-Y.}\ \bibnamefont {Huang}},
  \bibinfo {author} {\bibfnamefont {C.-H.}\ \bibnamefont {Lu}}, \bibinfo
  {author} {\bibfnamefont {L.}~\bibnamefont {Rego}}, \bibinfo {author}
  {\bibfnamefont {D.~D.}\ \bibnamefont {Hickstein}}, \bibinfo {author}
  {\bibfnamefont {J.~L.}\ \bibnamefont {Ellis}}, \bibinfo {author}
  {\bibfnamefont {A.}~\bibnamefont {Jaron-Becker}}, \bibinfo {author}
  {\bibfnamefont {A.}~\bibnamefont {Becker}}, \bibinfo {author} {\bibfnamefont
  {S.-D.}\ \bibnamefont {Yang}}, \bibinfo {author} {\bibfnamefont {C.~G.}\
  \bibnamefont {Durfee}}, \bibinfo {author} {\bibfnamefont {L.}~\bibnamefont
  {Plaja}}, \bibinfo {author} {\bibfnamefont {H.~C.}\ \bibnamefont {Kapteyn}},
  \bibinfo {author} {\bibfnamefont {M.~M.}\ \bibnamefont {Murnane}}, \bibinfo
  {author} {\bibfnamefont {A.~H.}\ \bibnamefont {Kung}}, \ and\ \bibinfo
  {author} {\bibfnamefont {M.-C.}\ \bibnamefont {Chen}},\ }\href {\doibase
  10.1038/s41566-018-0145-0} {\bibfield  {journal} {\bibinfo  {journal} {Nature
  Photonics}\ }\textbf {\bibinfo {volume} {12}},\ \bibinfo {pages} {349}
  (\bibinfo {year} {2018})}\BibitemShut {NoStop}%
\bibitem [{\citenamefont {Ellis}\ \emph {et~al.}(2018)\citenamefont {Ellis},
  \citenamefont {Dorney}, \citenamefont {Hickstein}, \citenamefont {Brooks},
  \citenamefont {Gentry}, \citenamefont {Hern\'{a}ndez-Garc\'{i}a},
  \citenamefont {Zusin}, \citenamefont {Shaw}, \citenamefont {Nguyen},
  \citenamefont {Mancuso}, \citenamefont {Jansen}, \citenamefont {Witte},
  \citenamefont {Kapteyn},\ and\ \citenamefont {Murnane}}]{NON-Ellis}%
  \BibitemOpen
  \bibfield  {author} {\bibinfo {author} {\bibfnamefont {J.~L.}\ \bibnamefont
  {Ellis}}, \bibinfo {author} {\bibfnamefont {K.~M.}\ \bibnamefont {Dorney}},
  \bibinfo {author} {\bibfnamefont {D.~D.}\ \bibnamefont {Hickstein}}, \bibinfo
  {author} {\bibfnamefont {N.~J.}\ \bibnamefont {Brooks}}, \bibinfo {author}
  {\bibfnamefont {C.}~\bibnamefont {Gentry}}, \bibinfo {author} {\bibfnamefont
  {C.}~\bibnamefont {Hern\'{a}ndez-Garc\'{i}a}}, \bibinfo {author}
  {\bibfnamefont {D.}~\bibnamefont {Zusin}}, \bibinfo {author} {\bibfnamefont
  {J.~M.}\ \bibnamefont {Shaw}}, \bibinfo {author} {\bibfnamefont {Q.~L.}\
  \bibnamefont {Nguyen}}, \bibinfo {author} {\bibfnamefont {C.~A.}\
  \bibnamefont {Mancuso}}, \bibinfo {author} {\bibfnamefont {G.~S.~M.}\
  \bibnamefont {Jansen}}, \bibinfo {author} {\bibfnamefont {S.}~\bibnamefont
  {Witte}}, \bibinfo {author} {\bibfnamefont {H.~C.}\ \bibnamefont {Kapteyn}},
  \ and\ \bibinfo {author} {\bibfnamefont {M.~M.}\ \bibnamefont {Murnane}},\
  }\href {\doibase 10.1364/OPTICA.5.000479} {\bibfield  {journal} {\bibinfo
  {journal} {Optica}\ }\textbf {\bibinfo {volume} {5}},\ \bibinfo {pages} {479}
  (\bibinfo {year} {2018})}\BibitemShut {NoStop}%
\bibitem [{\citenamefont {Han}\ \emph {et~al.}(2023)\citenamefont {Han},
  \citenamefont {Ji}, \citenamefont {Ueda},\ and\ \citenamefont
  {W\"{o}rner}}]{NON-Han}%
  \BibitemOpen
  \bibfield  {author} {\bibinfo {author} {\bibfnamefont {M.}~\bibnamefont
  {Han}}, \bibinfo {author} {\bibfnamefont {J.-B.}\ \bibnamefont {Ji}},
  \bibinfo {author} {\bibfnamefont {K.}~\bibnamefont {Ueda}}, \ and\ \bibinfo
  {author} {\bibfnamefont {H.~J.}\ \bibnamefont {W\"{o}rner}},\ }\href
  {\doibase 10.1364/OPTICA.492741} {\bibfield  {journal} {\bibinfo  {journal}
  {Optica}\ }\textbf {\bibinfo {volume} {10}},\ \bibinfo {pages} {1044}
  (\bibinfo {year} {2023})}\BibitemShut {NoStop}%
\bibitem [{\citenamefont {Li}\ \emph {et~al.}(2018)\citenamefont {Li},
  \citenamefont {Lan}, \citenamefont {He}, \citenamefont {Zhu}, \citenamefont
  {Chen},\ and\ \citenamefont {Lu}}]{PRLLL}%
  \BibitemOpen
  \bibfield  {author} {\bibinfo {author} {\bibfnamefont {L.}~\bibnamefont
  {Li}}, \bibinfo {author} {\bibfnamefont {P.}~\bibnamefont {Lan}}, \bibinfo
  {author} {\bibfnamefont {L.}~\bibnamefont {He}}, \bibinfo {author}
  {\bibfnamefont {X.}~\bibnamefont {Zhu}}, \bibinfo {author} {\bibfnamefont
  {J.}~\bibnamefont {Chen}}, \ and\ \bibinfo {author} {\bibfnamefont
  {P.}~\bibnamefont {Lu}},\ }\href {\doibase 10.1103/PhysRevLett.120.223203}
  {\bibfield  {journal} {\bibinfo  {journal} {Phys. Rev. Lett.}\ }\textbf
  {\bibinfo {volume} {120}},\ \bibinfo {pages} {223203} (\bibinfo {year}
  {2018})}\BibitemShut {NoStop}%
\bibitem [{\citenamefont {Zhai}\ \emph {et~al.}(2021)\citenamefont {Zhai},
  \citenamefont {Zhu}, \citenamefont {Long}, \citenamefont {Shao},
  \citenamefont {Zhang}, \citenamefont {He}, \citenamefont {Tang},
  \citenamefont {Li}, \citenamefont {Lan}, \citenamefont {Yu},\ and\
  \citenamefont {Lu}}]{mixedZhai}%
  \BibitemOpen
  \bibfield  {author} {\bibinfo {author} {\bibfnamefont {C.}~\bibnamefont
  {Zhai}}, \bibinfo {author} {\bibfnamefont {X.}~\bibnamefont {Zhu}}, \bibinfo
  {author} {\bibfnamefont {J.}~\bibnamefont {Long}}, \bibinfo {author}
  {\bibfnamefont {R.}~\bibnamefont {Shao}}, \bibinfo {author} {\bibfnamefont
  {Y.}~\bibnamefont {Zhang}}, \bibinfo {author} {\bibfnamefont
  {L.}~\bibnamefont {He}}, \bibinfo {author} {\bibfnamefont {Q.}~\bibnamefont
  {Tang}}, \bibinfo {author} {\bibfnamefont {Y.}~\bibnamefont {Li}}, \bibinfo
  {author} {\bibfnamefont {P.}~\bibnamefont {Lan}}, \bibinfo {author}
  {\bibfnamefont {B.}~\bibnamefont {Yu}}, \ and\ \bibinfo {author}
  {\bibfnamefont {P.}~\bibnamefont {Lu}},\ }\href {\doibase
  10.1103/PhysRevA.103.033114} {\bibfield  {journal} {\bibinfo  {journal}
  {Phys. Rev. A}\ }\textbf {\bibinfo {volume} {103}},\ \bibinfo {pages}
  {033114} (\bibinfo {year} {2021})}\BibitemShut {NoStop}%
\bibitem [{\citenamefont {{M. J. Frisch \it{et al.}}}()}]{g09}%
  \BibitemOpen
  \bibfield  {author} {\bibinfo {author} {\bibnamefont {{M. J. Frisch \it{et
  al.}}}},\ }\href@noop {} {\enquote {\bibinfo {title} {{Gaussian 09, Revision
  A.02}},}\ }\bibinfo {note} {Gaussian, Inc. Wallingford CT, 2016}\BibitemShut
  {NoStop}%
\bibitem [{\citenamefont {Lewenstein}\ \emph {et~al.}(1994)\citenamefont
  {Lewenstein}, \citenamefont {Balcou}, \citenamefont {Ivanov}, \citenamefont
  {L'Huillier},\ and\ \citenamefont {Corkum}}]{SFA}%
  \BibitemOpen
  \bibfield  {author} {\bibinfo {author} {\bibfnamefont {M.}~\bibnamefont
  {Lewenstein}}, \bibinfo {author} {\bibfnamefont {P.}~\bibnamefont {Balcou}},
  \bibinfo {author} {\bibfnamefont {M.~Y.}\ \bibnamefont {Ivanov}}, \bibinfo
  {author} {\bibfnamefont {A.}~\bibnamefont {L'Huillier}}, \ and\ \bibinfo
  {author} {\bibfnamefont {P.~B.}\ \bibnamefont {Corkum}},\ }\href {\doibase
  10.1103/PhysRevA.49.2117} {\bibfield  {journal} {\bibinfo  {journal} {Phys.
  Rev. A}\ }\textbf {\bibinfo {volume} {49}},\ \bibinfo {pages} {2117}
  (\bibinfo {year} {1994})}\BibitemShut {NoStop}%
\bibitem [{\citenamefont {Neufeld}\ \emph {et~al.}(2019)\citenamefont
  {Neufeld}, \citenamefont {Podolsky},\ and\ \citenamefont {Cohen}}]{OCSel}%
  \BibitemOpen
  \bibfield  {author} {\bibinfo {author} {\bibfnamefont {O.}~\bibnamefont
  {Neufeld}}, \bibinfo {author} {\bibfnamefont {D.}~\bibnamefont {Podolsky}}, \
  and\ \bibinfo {author} {\bibfnamefont {O.}~\bibnamefont {Cohen}},\ }\href
  {\doibase 10.1038/s41467-018-07935-y} {\bibfield  {journal} {\bibinfo
  {journal} {Nature Communications}\ }\textbf {\bibinfo {volume} {10}},\
  \bibinfo {pages} {405} (\bibinfo {year} {2019})}\BibitemShut {NoStop}%
\bibitem [{\citenamefont {Liu}\ \emph {et~al.}(2016)\citenamefont {Liu},
  \citenamefont {Zhu}, \citenamefont {Li}, \citenamefont {Li}, \citenamefont
  {Zhang}, \citenamefont {Lan},\ and\ \citenamefont {Lu}}]{liuxiSel}%
  \BibitemOpen
  \bibfield  {author} {\bibinfo {author} {\bibfnamefont {X.}~\bibnamefont
  {Liu}}, \bibinfo {author} {\bibfnamefont {X.}~\bibnamefont {Zhu}}, \bibinfo
  {author} {\bibfnamefont {L.}~\bibnamefont {Li}}, \bibinfo {author}
  {\bibfnamefont {Y.}~\bibnamefont {Li}}, \bibinfo {author} {\bibfnamefont
  {Q.}~\bibnamefont {Zhang}}, \bibinfo {author} {\bibfnamefont
  {P.}~\bibnamefont {Lan}}, \ and\ \bibinfo {author} {\bibfnamefont
  {P.}~\bibnamefont {Lu}},\ }\href {\doibase 10.1103/PhysRevA.94.033410}
  {\bibfield  {journal} {\bibinfo  {journal} {Phys. Rev. A}\ }\textbf {\bibinfo
  {volume} {94}},\ \bibinfo {pages} {033410} (\bibinfo {year}
  {2016})}\BibitemShut {NoStop}%
\bibitem [{\citenamefont {McFarland}\ \emph {et~al.}(2009)\citenamefont
  {McFarland}, \citenamefont {Farrell}, \citenamefont {Bucksbaum},\ and\
  \citenamefont {G\"uhr}}]{mixed2009}%
  \BibitemOpen
  \bibfield  {author} {\bibinfo {author} {\bibfnamefont {B.~K.}\ \bibnamefont
  {McFarland}}, \bibinfo {author} {\bibfnamefont {J.~P.}\ \bibnamefont
  {Farrell}}, \bibinfo {author} {\bibfnamefont {P.~H.}\ \bibnamefont
  {Bucksbaum}}, \ and\ \bibinfo {author} {\bibfnamefont {M.}~\bibnamefont
  {G\"uhr}},\ }\href {\doibase 10.1103/PhysRevA.80.033412} {\bibfield
  {journal} {\bibinfo  {journal} {Phys. Rev. A}\ }\textbf {\bibinfo {volume}
  {80}},\ \bibinfo {pages} {033412} (\bibinfo {year} {2009})}\BibitemShut
  {NoStop}%
\bibitem [{\citenamefont {Bertrand}\ \emph {et~al.}(2013)\citenamefont
  {Bertrand}, \citenamefont {Wörner}, \citenamefont {Salières}, \citenamefont
  {Villeneuve},\ and\ \citenamefont {Corkum}}]{NPlink}%
  \BibitemOpen
  \bibfield  {author} {\bibinfo {author} {\bibfnamefont {J.~B.}\ \bibnamefont
  {Bertrand}}, \bibinfo {author} {\bibfnamefont {H.~J.}\ \bibnamefont
  {Wörner}}, \bibinfo {author} {\bibfnamefont {P.}~\bibnamefont {Salières}},
  \bibinfo {author} {\bibfnamefont {D.~M.}\ \bibnamefont {Villeneuve}}, \ and\
  \bibinfo {author} {\bibfnamefont {P.~B.}\ \bibnamefont {Corkum}},\ }\href
  {\doibase 10.1038/nphys2540} {\bibfield  {journal} {\bibinfo  {journal}
  {Nature Physics}\ }\textbf {\bibinfo {volume} {9}},\ \bibinfo {pages} {174}
  (\bibinfo {year} {2013})}\BibitemShut {NoStop}%
\bibitem [{\citenamefont {Smirnova}\ \emph {et~al.}(2009)\citenamefont
  {Smirnova}, \citenamefont {Mairesse}, \citenamefont {Patchkovskii},
  \citenamefont {Dudovich}, \citenamefont {Villeneuve}, \citenamefont
  {Corkum},\ and\ \citenamefont {Ivanov}}]{rpnature}%
  \BibitemOpen
  \bibfield  {author} {\bibinfo {author} {\bibfnamefont {O.}~\bibnamefont
  {Smirnova}}, \bibinfo {author} {\bibfnamefont {Y.}~\bibnamefont {Mairesse}},
  \bibinfo {author} {\bibfnamefont {S.}~\bibnamefont {Patchkovskii}}, \bibinfo
  {author} {\bibfnamefont {N.}~\bibnamefont {Dudovich}}, \bibinfo {author}
  {\bibfnamefont {D.}~\bibnamefont {Villeneuve}}, \bibinfo {author}
  {\bibfnamefont {P.}~\bibnamefont {Corkum}}, \ and\ \bibinfo {author}
  {\bibfnamefont {M.~Y.}\ \bibnamefont {Ivanov}},\ }\href {\doibase
  10.1038/nature08253} {\bibfield  {journal} {\bibinfo  {journal} {Nature}\
  }\textbf {\bibinfo {volume} {460}},\ \bibinfo {pages} {972} (\bibinfo {year}
  {2009})}\BibitemShut {NoStop}%
\bibitem [{\citenamefont {Camper}\ \emph {et~al.}(2023)\citenamefont {Camper},
  \citenamefont {Ferr\'e}, \citenamefont {Blanchet}, \citenamefont {Descamps},
  \citenamefont {Lin}, \citenamefont {Petit}, \citenamefont {Lucchese},
  \citenamefont {Sali\`eres}, \citenamefont {Ruchon},\ and\ \citenamefont
  {Mairesse}}]{rpprl}%
  \BibitemOpen
  \bibfield  {author} {\bibinfo {author} {\bibfnamefont {A.}~\bibnamefont
  {Camper}}, \bibinfo {author} {\bibfnamefont {A.}~\bibnamefont {Ferr\'e}},
  \bibinfo {author} {\bibfnamefont {V.}~\bibnamefont {Blanchet}}, \bibinfo
  {author} {\bibfnamefont {D.}~\bibnamefont {Descamps}}, \bibinfo {author}
  {\bibfnamefont {N.}~\bibnamefont {Lin}}, \bibinfo {author} {\bibfnamefont
  {S.}~\bibnamefont {Petit}}, \bibinfo {author} {\bibfnamefont
  {R.}~\bibnamefont {Lucchese}}, \bibinfo {author} {\bibfnamefont
  {P.}~\bibnamefont {Sali\`eres}}, \bibinfo {author} {\bibfnamefont
  {T.}~\bibnamefont {Ruchon}}, \ and\ \bibinfo {author} {\bibfnamefont
  {Y.}~\bibnamefont {Mairesse}},\ }\href {\doibase
  10.1103/PhysRevLett.130.083201} {\bibfield  {journal} {\bibinfo  {journal}
  {Phys. Rev. Lett.}\ }\textbf {\bibinfo {volume} {130}},\ \bibinfo {pages}
  {083201} (\bibinfo {year} {2023})}\BibitemShut {NoStop}%
\bibitem [{\citenamefont {Salières}\ \emph {et~al.}(2001)\citenamefont
  {Salières}, \citenamefont {Carré}, \citenamefont {Déroff}, \citenamefont
  {Grasbon}, \citenamefont {Paulus}, \citenamefont {Walther}, \citenamefont
  {Kopold}, \citenamefont {Becker}, \citenamefont {Milošević}, \citenamefont
  {Sanpera},\ and\ \citenamefont {Lewenstein}}]{QO}%
  \BibitemOpen
  \bibfield  {author} {\bibinfo {author} {\bibfnamefont {P.}~\bibnamefont
  {Salières}}, \bibinfo {author} {\bibfnamefont {B.}~\bibnamefont {Carré}},
  \bibinfo {author} {\bibfnamefont {L.~L.}\ \bibnamefont {Déroff}}, \bibinfo
  {author} {\bibfnamefont {F.}~\bibnamefont {Grasbon}}, \bibinfo {author}
  {\bibfnamefont {G.~G.}\ \bibnamefont {Paulus}}, \bibinfo {author}
  {\bibfnamefont {H.}~\bibnamefont {Walther}}, \bibinfo {author} {\bibfnamefont
  {R.}~\bibnamefont {Kopold}}, \bibinfo {author} {\bibfnamefont
  {W.}~\bibnamefont {Becker}}, \bibinfo {author} {\bibfnamefont {D.~B.}\
  \bibnamefont {Milošević}}, \bibinfo {author} {\bibfnamefont
  {A.}~\bibnamefont {Sanpera}}, \ and\ \bibinfo {author} {\bibfnamefont
  {M.}~\bibnamefont {Lewenstein}},\ }\href {\doibase 10.1126/science.108836}
  {\bibfield  {journal} {\bibinfo  {journal} {Science}\ }\textbf {\bibinfo
  {volume} {292}},\ \bibinfo {pages} {902} (\bibinfo {year} {2001})},\ \Eprint
  {http://arxiv.org/abs/https://www.science.org/doi/pdf/10.1126/science.108836}
  {https://www.science.org/doi/pdf/10.1126/science.108836} \BibitemShut
  {NoStop}%
\bibitem [{\citenamefont {Baykusheva}\ \emph {et~al.}(2017)\citenamefont
  {Baykusheva}, \citenamefont {Brennecke}, \citenamefont {Lein},\ and\
  \citenamefont {W\"orner}}]{saddlePRL}%
  \BibitemOpen
  \bibfield  {author} {\bibinfo {author} {\bibfnamefont {D.}~\bibnamefont
  {Baykusheva}}, \bibinfo {author} {\bibfnamefont {S.}~\bibnamefont
  {Brennecke}}, \bibinfo {author} {\bibfnamefont {M.}~\bibnamefont {Lein}}, \
  and\ \bibinfo {author} {\bibfnamefont {H.~J.}\ \bibnamefont {W\"orner}},\
  }\href {\doibase 10.1103/PhysRevLett.119.203201} {\bibfield  {journal}
  {\bibinfo  {journal} {Phys. Rev. Lett.}\ }\textbf {\bibinfo {volume} {119}},\
  \bibinfo {pages} {203201} (\bibinfo {year} {2017})}\BibitemShut {NoStop}%
\bibitem [{\citenamefont {Dorney}\ \emph {et~al.}(2021)\citenamefont {Dorney},
  \citenamefont {Fan}, \citenamefont {Nguyen}, \citenamefont {Ellis},
  \citenamefont {Hickstein}, \citenamefont {Brooks}, \citenamefont {Zusin},
  \citenamefont {Gentry}, \citenamefont {Hern\'{a}ndez-Garc\'{i}a},
  \citenamefont {Kapteyn},\ and\ \citenamefont {Murnane}}]{scOE}%
  \BibitemOpen
  \bibfield  {author} {\bibinfo {author} {\bibfnamefont {K.~M.}\ \bibnamefont
  {Dorney}}, \bibinfo {author} {\bibfnamefont {T.}~\bibnamefont {Fan}},
  \bibinfo {author} {\bibfnamefont {Q.~L.~D.}\ \bibnamefont {Nguyen}}, \bibinfo
  {author} {\bibfnamefont {J.~L.}\ \bibnamefont {Ellis}}, \bibinfo {author}
  {\bibfnamefont {D.~D.}\ \bibnamefont {Hickstein}}, \bibinfo {author}
  {\bibfnamefont {N.}~\bibnamefont {Brooks}}, \bibinfo {author} {\bibfnamefont
  {D.}~\bibnamefont {Zusin}}, \bibinfo {author} {\bibfnamefont
  {C.}~\bibnamefont {Gentry}}, \bibinfo {author} {\bibfnamefont
  {C.}~\bibnamefont {Hern\'{a}ndez-Garc\'{i}a}}, \bibinfo {author}
  {\bibfnamefont {H.~C.}\ \bibnamefont {Kapteyn}}, \ and\ \bibinfo {author}
  {\bibfnamefont {M.~M.}\ \bibnamefont {Murnane}},\ }\href {\doibase
  10.1364/OE.440813} {\bibfield  {journal} {\bibinfo  {journal} {Opt. Express}\
  }\textbf {\bibinfo {volume} {29}},\ \bibinfo {pages} {38119} (\bibinfo {year}
  {2021})}\BibitemShut {NoStop}%
\end{thebibliography}%
\end{document}